\makeatletter
\let\old@footnotetext=\@footnotetext
\makeatother
\documentclass{aastex}
\makeatletter
\let\@footnotetext=\old@footnotetext
\makeatother
\usepackage{emulateapj5}

\newcommand{\cxo}{{\it Chandra X-ray Observatory}}
\newcommand{\chandra}{{\it Chandra}}

\newcommand{\asca}{{\it ASCA}}

\newcommand{\co}{CXOU~J050552.3$-$680141}
\newcommand{\snr}{SNR 0506$-$68.0}
\newcommand{\lsim}{\hbox{\raise.35ex\rlap{$<$}\lower.6ex\hbox{$\sim$}\ }}
\newcommand{\gsim}{\hbox{\raise.35ex\rlap{$>$}\lower.6ex\hbox{$\sim$}\ }}

\slugcomment{}

\submitted{Received 2006 April 5; accepted 2006 May 31}
\accepted{To appear in the Astrophysical Journal Letters}

\shorttitle{\chandra\ View of SNR 0506$-$68.0}
\shortauthors{Hughes et al.}

\begin{document}

\title{The {\it Chandra} View of the Supernova 
Remnant 0506$-$68.0 in the Large Magellanic Cloud}

\author{
 John P.~Hughes\altaffilmark{1}, 
 Marc Rafelski\altaffilmark{2},
 Jessica S.~Warren\altaffilmark{1}, 
 Cara Rakowski\altaffilmark{3},
 Patrick Slane\altaffilmark{3},
 David Burrows\altaffilmark{4},
 and
 John Nousek\altaffilmark{4}
}
\altaffiltext{1}{Department of Physics and Astronomy, Rutgers
University, 136 Frelinghuysen Road, Piscataway, NJ 08854-8019;
jph@physics.rutgers.edu, jesawyer@physics.rutgers.edu}
\altaffiltext{2}{UCLA, Physics and Astronomy Building, 430 Portola 
Plaza, Box 951547, Los Angeles, CA 90095-1547; marcar@astro.ucla.edu}
\altaffiltext{3}{Harvard-Smithsonian Center for Astrophysics, 60 Garden 
Street,Cambridge, MA 02138; rakowski@head-cfa.harvard.edu, 
slane@head-cfa.harvard.edu}
\altaffiltext{4}{Department of Astronomy and Astrophysics, Pennsylvania State
University, 525 Davey Laboratory, University Park, PA. 16802;
dxb15@psu.edu, nousek@astro.psu.edu}

\begin{abstract}

A new \chandra\ observation of \snr\ (also called N23) reveals a
complex, highly structured morphology in the low energy X-ray band and
an isolated compact central object in the high energy band.  Spectral
analysis indicates that the X-ray emission overall is dominated by
thermal gas whose composition is consistent with swept-up ambient
material.  There is a strong gradient in ambient density across the
diameter of the remnant. Toward the southeast, near a prominent star
cluster, the emitting density is 10--23 cm$^{-3}$, while toward the
northwest it has dropped to a value of only 1 cm$^{-3}$.  The total
extent of the X-ray remnant is $100^{\prime\prime} \times
120^{\prime\prime}$ (24 pc $\times$ 29 pc for a distance of 50 kpc),
somewhat larger than previously known. The remnant's age is estimated
to be $\sim$4600 yr.  One part of the remnant shows evidence for
enhanced O, Ne, and perhaps Mg abundances, which is interpreted as
evidence for ejecta from a massive star core collapse supernova. The
compact central object has a luminosity of a few times $10^{33}$ ergs
s$^{-1}$ and no obvious radio or optical counterpart. It does not show
an extended nebula or pulsed emission as expected from a young
energetic pulsar, but resembles the compact central objects seen in
other core collapse SNe, such as Cas A.

\end{abstract}

\keywords{
ISM: individual (LHA 120-N 23, \snr) --- 
shock waves ---
supernova remnants ---
X-rays: individual (\co) ---
X-rays: ISM
}

\section{Introduction}

The \cxo\ has been a revolutionary tool for studying supernova
remnants (SNRs) in the Large Magellanic Cloud (LMC).  The telescope's
superb spatial resolution has revealed newly processed metals in the
interiors of most LMC SNRs and at least one new pulsar wind nebula
\citep[see][and reference therein]{weishugh05}.  Studying the full
sample of LMC SNRs is essential to gain a more complete understanding
of the evolution of metal-rich ejecta from its synthesis in stars and
SNe to its integration into the interstellar medium.

Although relatively bright in the radio and X-ray bands \citep[it is
ranked ninth brightest in the soft X-ray band and eleventh brightest
at 408 MHz;][]{math83}, \snr\ (also called N23) has not been
extensively studied.  \citet{chuken88} suggest that it had a massive
star progenitor due to the high density of OB stars in the environment
and its close proximity to \ion{H}{2} regions and star clusters.
There is a strong gradient in brightness across the remnant in the
X-ray \citep{math83} and radio \citep{dick98} bands.  Its integrated
X-ray emission \citep{hughes98} is dominated by soft thermal emission
from swept-up ambient medium with the lowered abundances ($\sim$0.3
solar) typical of the LMC.

According to \citet{math83}, \snr\ is an intermediate sized remnant
($\sim$11 pc in diameter).  All smaller LMC remnants show evidence for
youth (i.e., an energetic pulsar or ejecta-dominated emission), while
nearly all larger remnants are dominated by swept-up ambient medium
\citep{hughes98}.  Of course, size alone is an imperfect predictor of
evolutionary state, yet the possibility that \snr\ is in a
transitional stage of evolution (between ejecta-dominated and Sedov)
makes it worth studying. The original goals of our observation were to
find and categorize X-ray--emitting SN ejecta, search for a
pulsar-powered synchrotron nebula or other compact X-ray star, and
clarify the evolutionary state of \snr\ through spatially resolved
X-ray spectroscopy to better understand the nature of its progenitor
star and its effects on the local environment.  While this work was
being completed, an independent report \citep{haya1,haya2} noting a
spectrally hard, X-ray compact object in N23 came to our attention.
In the following we present the results of our comprehensive
exposition of the \chandra\ data in which we arrive at a different
conclusion regarding the nature of the compact star. Throughout, we
assume a distance of 50 kpc to the LMC.

\section{Data Reduction and Analysis}

We observed \snr\ with \chandra\ on 2002 Dec 29 using the ACIS-S
detector in full-frame timed exposure mode (ObsId 2762). Standard
cleaning procedures were applied to the data including filtering for
times of high background that resulted in a useful live-time corrected
exposure of 34744 s.  Event pulse heights were corrected for time
dependent gain effects and spectral response functions accounted for
absorption due to contamination build-up on the ACIS filters. The
optical field of the SNR is quite crowded and it was not possible to
identify unambiguous counterparts to serendipitous \chandra\ sources
and thereby improve the astrometric accuracy of the X-ray
data. Nevertheless the observation is not affected by any known aspect
offset and therefore we expect the absolute positional uncertainty to
be 0.6$^{\prime\prime}$ (radius, at 90\% confidence).

\begin{center}
\begin{minipage}[h]{0.47\textwidth} 
\epsfxsize=3.0truein\epsfbox[230 56 553 707]{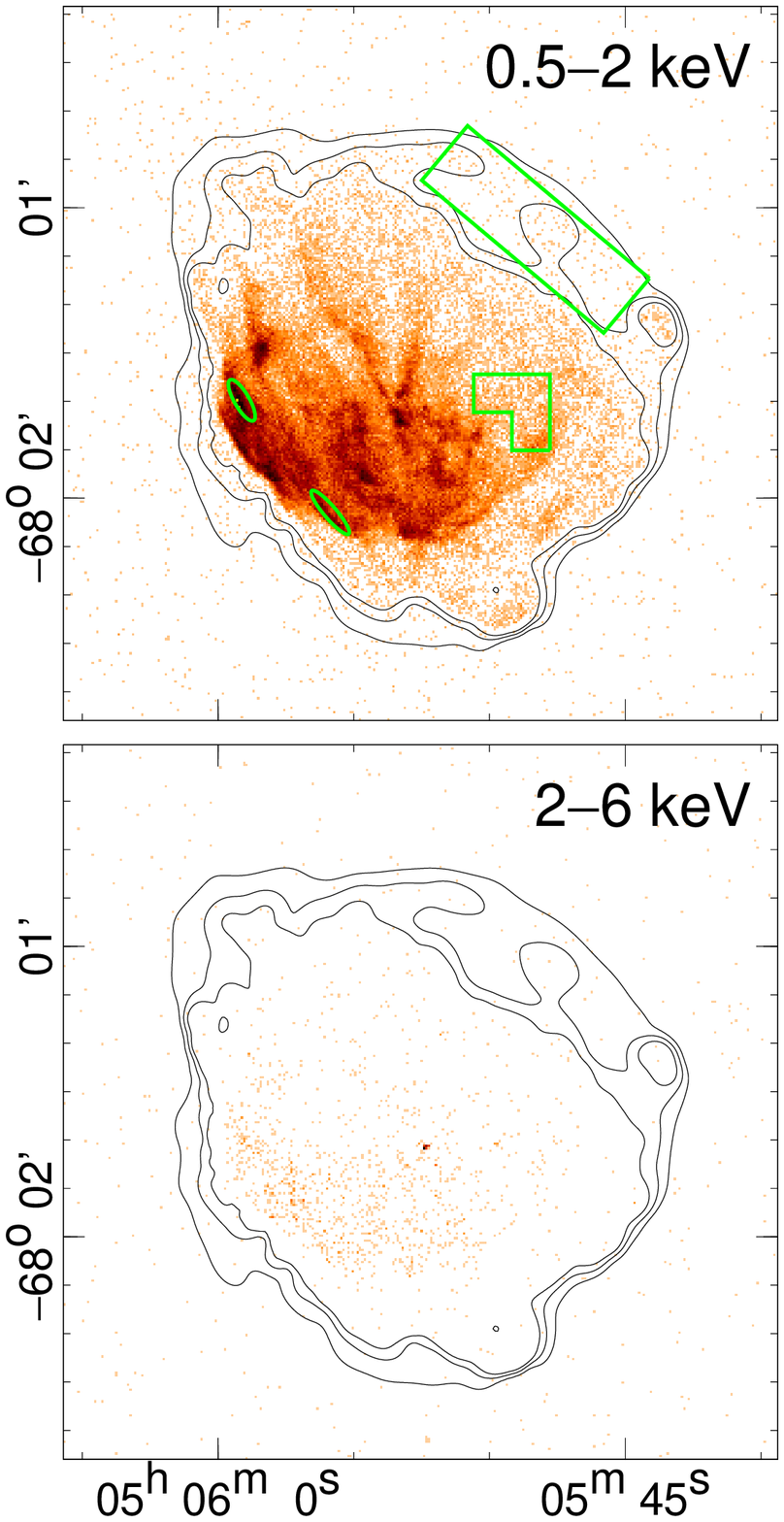}
\vspace{-0.25truein}
\figcaption{Soft (top) and hard (bottom) band \chandra\ X-ray images of \snr. 
Spectral extraction regions are shown in green in the top panel.  Contours 
delineate the faintest outer extent of the remnant from a smoothed image.
\label{xray}}
\end{minipage}
\end{center}

\subsection{Gaseous Remnant}

Figure~\ref{xray} shows \chandra\ images of \snr\ in two broad energy
bands (0.5--2 keV and 2--6 keV) along with contours indicating the
faint outermost extent of the remnant from an adaptively smoothed
image (not shown).  Nearly all of the X-ray emission occurs below 2
keV except for a central compact object (see below) and some of the
brightest portions of the limb. Those bright portions lie mostly
toward the east and southeast and appear highly structured and
filamentary. The northwestern half of the remnant is much fainter and
shows a hint of limb-brightening. The overall extent of the remnant is
$100^{\prime\prime} \times 120^{\prime\prime}$, larger than previous
estimates of the X-ray size \citep{math83,will99}. The integrated
X-ray spectrum (Fig.~\ref{spectrum}) clearly indicates the
characteristics of soft thermal emission and closely resembles the
\asca\ spectrum \citep{hughes98}.

{\parindent=0cm 
\begin{small}
%\begin{minipage}[t]{83mm}
\begin{minipage}[t]{0.47\textwidth}
\begin{center}
{\noindent{TABLE 1}}\\
{\noindent{\sc Spectral Model Fit Results for SNR Filaments}}\\[6pt]
\begin{tabular}{@{}llcc@{}}
\hline\hline\\[-4pt]
 & \multicolumn{2}{c}{Value and Uncertainty (1 $\sigma$)}\\
Parameter & East rim & Southeast rim \\[4pt]
\hline\\[-4pt]
$N_{\rm H}({\rm LMC})$ (H atoms cm$^{-2}$) & $3.3^{+0.10}_{-0.07} \times 10^{21}$   & $2.1^{+0.4}_{-0.1} \times 10^{21}$ \\
$kT$ (keV)                          & $0.48\pm0.14$   & $0.77\pm0.4$ \\
$\log_{10}(n_et/{\rm cm^{-3}\,s}$)  & $10.6^{+0.5}_{-0.3}$  & $11.2\pm 0.3$ \\
Oxygen                              & $0.15\pm0.05$  & $0.30\pm0.18$ \\
Neon                                & $0.23\pm0.09$  & $0.32\pm0.09$ \\
Magnesium                           & $0.18\pm0.13$  & $0.21\pm0.17$ \\
Iron                                & $0.13\pm0.06$  & $0.20\pm0.15$ \\
$n_e n_{\rm H} V/(4 \pi D^2)$ (cm$^{-5}$) & $9.75\times 10^{10}$ 
   & $2.28\times 10^{10}$ \\
$\chi^2$/d.o.f                 &  94.8/56 &  42.4/49 \\
\hline\\[-8pt]
\end{tabular}
\end{center}
\label{tab_snr}
\end{minipage}
\end{small}
}

Two representative spectra from portions of the bright eastern rim
(hereafter east rim and southeast rim, indicated by the two ellipses
on Fig.~\ref{xray}) were extracted, regrouped to contain at least 20
counts per channel, and analyzed with planar-shock nonequilibrium
ionization (NEI) models \citep{hughes00}.  The fits included Galactic
absorption with solar composition and column density fixed to the
value $5.8\times 10^{20}$ atoms cm$^{-2}$ from \citet{dilo90}
and LMC absorption with variable $N_{\rm H}({\rm LMC})$ and metal
abundances of 0.3 times solar.  Numerical results are given in
Table~\ref{tab_snr} and the spectral data with best-fitted models are
plotted in Fig.~\ref{spectrum} (as blue and green curves).  Only the
abundances of O, Ne, Mg, and Fe in the emission model were allowed to
vary freely, all other species had their abundances fixed to 0.3 times
solar.  The derived abundances clearly show that the bright rim
emission is dominated by gas with LMC composition.  We also extracted
the spectrum of the faint western limb as well (from the large
rectangular region to the west of Fig.~\ref{xray}).  All elemental
abundances were fixed to 0.3 times solar, because of the low
statistics of this spectrum ($\sim$220 counts above background).
Furthermore the uncertainties on the fitted thermodynamic quantities
are large, so we only quote best fit values as indicators of the
thermodynamic state: $N_{\rm H}{(\rm LMC})\sim 7 \times 10^{21}$ H
atoms cm$^{-2}$, $kT\sim0.3$ keV, $\log_{10}(n_et/{\rm
cm^{-3}\,s})\sim 9.8$, and $n_e n_{\rm H} V/(4 \pi D^2)\sim 4.5\times
10^{10}$ cm$^{-5}$.  The fit was statistically acceptable ($\chi^2 =
7.2$ for 8 degrees of freedom).

Numerous spectra were extracted from elsewhere across the remnant
including both filament and inter-filament regions, but in nearly all
cases the gas composition was consistent with LMC abundances.  Only
one region showed any evidence for elevated abundances (the inverted
``L''-shaped region in Fig.~\ref{xray}). This spectrum's relatively
strong emission around 0.6 keV (see cyan curve in Fig.~\ref{spectrum})
clearly sets it apart from the rim spectra discussed above.  Our
spectral fit prefers high temperatures ($>$0.7 keV) and enhanced
abundances for O ($>$0.7) and Ne ($>$0.8), and perhaps Mg
($>$0.3). Note that these values represent the 90\% confidence lower
bound on the parameters. Over this range the Fe abundance is clearly
constrained to be at or below the LMC value.  For temperatures lower
than 0.7 keV, the abundance ratio of O to Fe (relative to solar) grows
even larger and at 0.5 keV the ratio is $>$20. Although the precise
values are somewhat indeterminant, we conclude with some confidence
that the low Z elements are enhanced in this particular part of the
remnant.

\subsection{Compact Star}

Close to the geometric center of the SNR an isolated X-ray point
source (hereafter \co) is clearly seen in the 2--6 keV band
(Fig.~\ref{xray}) at a position of RA=05:05:52.32, Decl.=$-$68:01:41.2
(epoch J2000). The detection in this hard X-ray band is significant;
we obtain 31 net counts within a 2 pixel radius with an expectation
value from background alone of 1.4 counts. The source is also detected
in the soft band (0.5--2 keV) at approximately the same significance
level.  The net (local background subtracted) count rate of the source
in the 0.5-6 keV band is $(2.8 \pm 0.4)\times 10^{-3}$ s$^{-1}$.

\begin{center}
\begin{minipage}[t]{0.47\textwidth} 
\epsfxsize=3.5truein\epsfbox{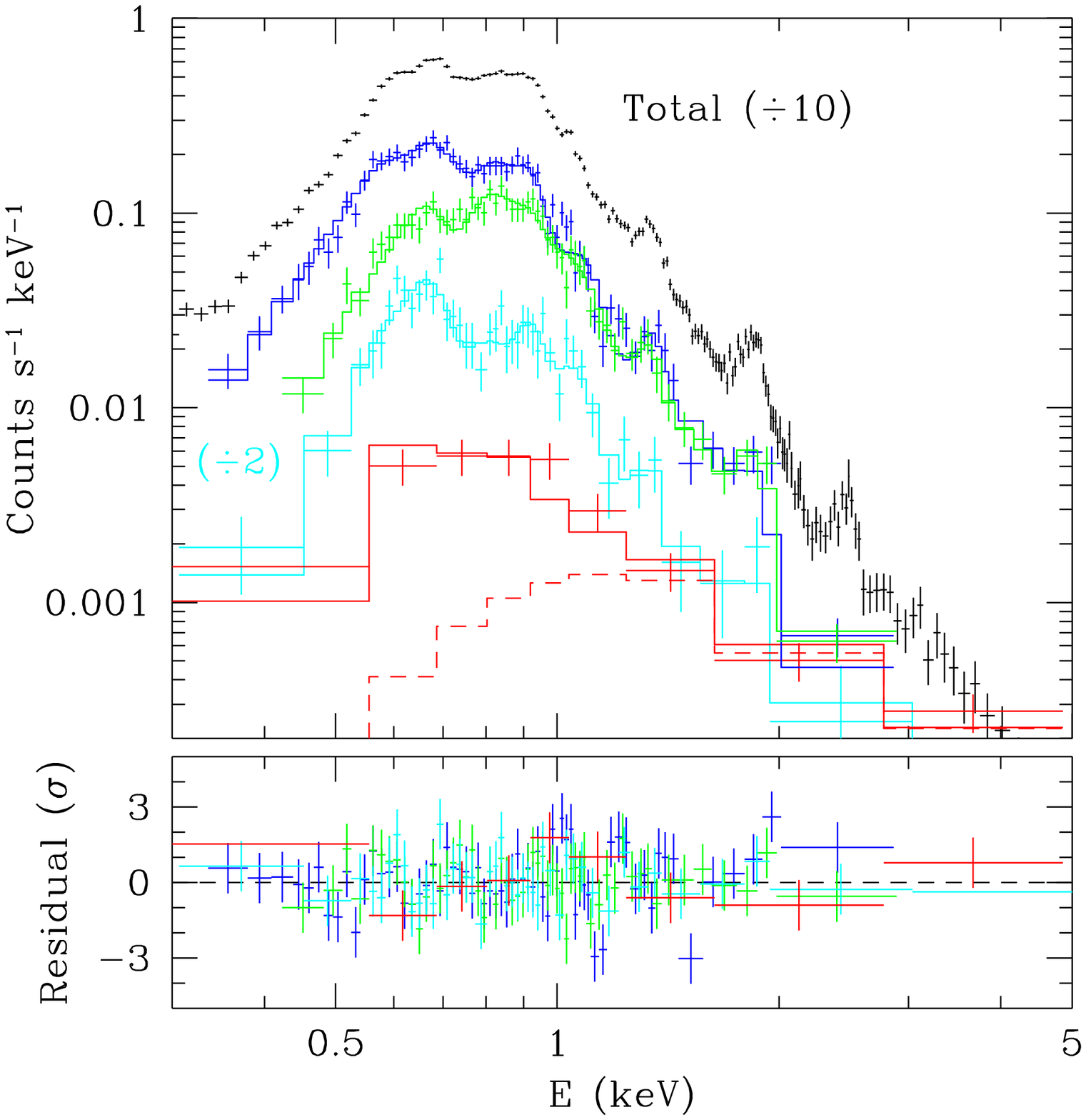}
\figcaption{Spectra and best fit models for several positions in \snr. In
order from the top the spectra are from the entire remnant (black), the
east bright rim (blue), southeast rim (green), enhanced abundance
region (cyan), and the central compact object (red).
\label{spectrum}}
\end{minipage}
\end{center}

We simulated the \chandra\ point-spread-function (PSF) at the observed
position of the compact object using \chandra\ Ray Tracer
(ChaRT).\footnote{ChaRT is available on-line at:
http://asc.harvard.edu/chart/index.html} The observed structure and
radial profile of the star are entirely consistent with the PSF and we
find no evidence for extended emission (for a close-up view of the
point source see the insert box in Fig.~\ref{opt}).

{\parindent=0cm 
\begin{small}
%\begin{minipage}[t]{83mm}
\begin{minipage}[t]{0.47\textwidth}
\begin{center}
{\noindent{TABLE 2}}\\
{\noindent{\sc Spectral Model Fit Results for Compact Source}}\\[6pt]
\begin{tabular}{@{}lcc@{}}
\hline\hline\\[-4pt]
          & Value and \\[2pt]
Parameter & Uncertainty (1 $\sigma$)\\[4pt]
\hline\\[-4pt]
\multicolumn{2}{c}{Power Law Model}\\
$N_{\rm H}({\rm LMC})$ (H atoms cm$^{-2}$) & $3.7^{+2.0}_{-2.4} 
                                                 \times 10^{21}$ \\
$\alpha_{\rm P}$                & $1.5^{+0.7}_{-0.6}$ \\
$F_{\rm E}(1\, \rm keV)$ (photon s$^{-1}$ cm$^{-2}$ keV$^{-1}$)
  & $4.5^{+2.9}_{-2.1} \times 10^{-6}$ \\
$\chi^2$/d.o.f           &  10.0/5 \\
\multicolumn{2}{c}{Blackbody Model}\\
$N_{\rm H}({\rm LMC})$ (H atoms cm$^{-2}$) & $5.7^{+1.5}_{-1.4} 
                                                \times 10^{21}$ \\
$kT$ (keV)                      & $0.96^{+0.36}_{-0.20}$ \\
Norm $(R/1\,{\rm km})^2/(D/10\,{\rm kpc})^2$  
    & $3.5^{+1.6}_{-2.6} \times 10^{-3}$ \\
$\chi^2$/d.o.f           &  12.9/5 \\
\hline\\[-8pt]
\end{tabular}
\end{center}
\label{tab_star}
\end{minipage}
\end{small}
}

\begin{center}
\begin{minipage}[t]{0.47\textwidth} 
\epsfxsize=3.5truein\epsfbox{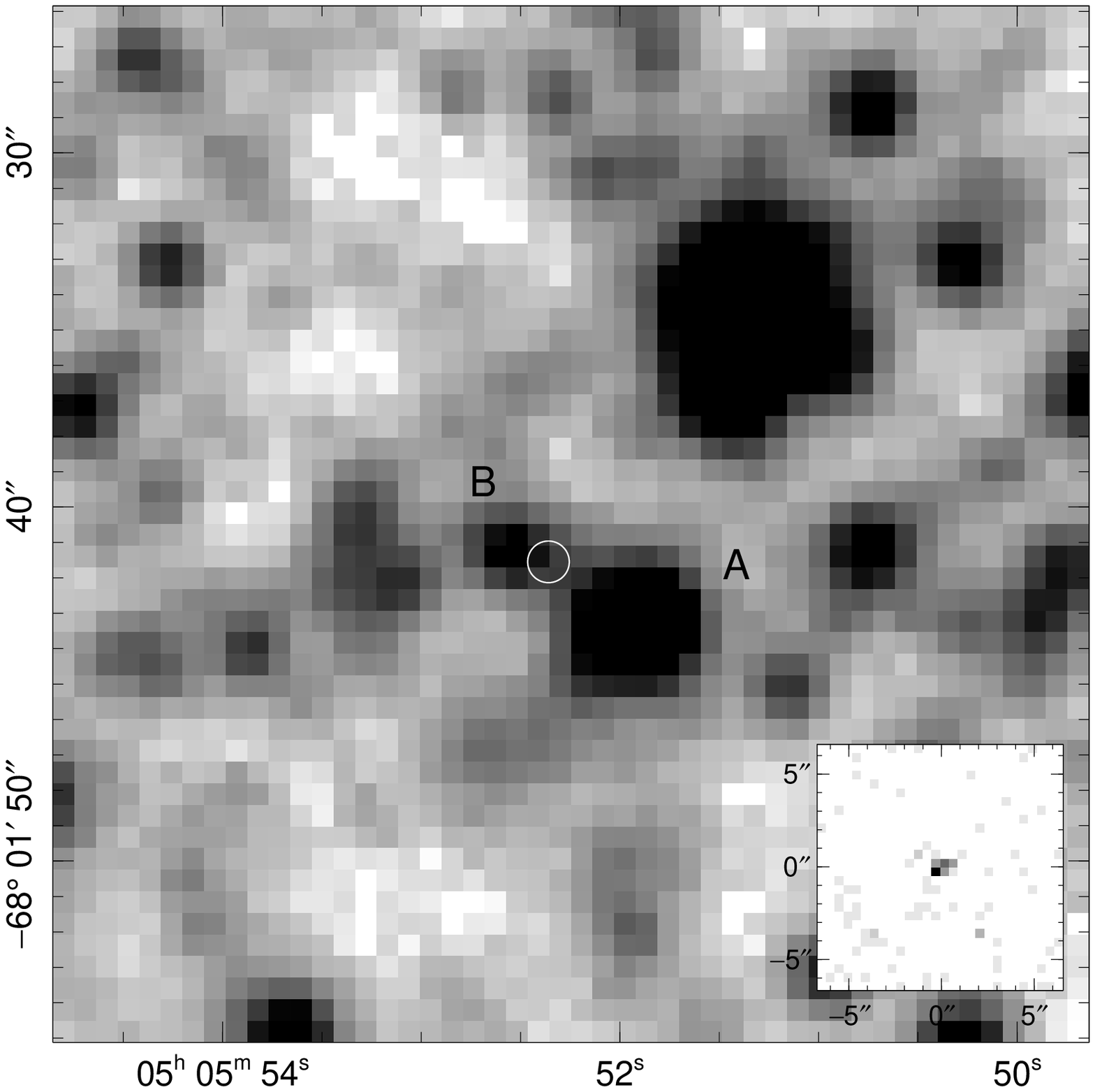}
\vspace{-0.25truein}
\figcaption{Broad band blue optical image from the MACHO project in the
vicinity of the compact central X-ray source.  Stars labeled A and B
are closest to the X-ray source, whose error circle is shown in white.
The X-ray data in the 2--6 keV band are shown in the small insert
figure to the lower right. The slight E-W extent of the X-ray source
is entirely due to the \chandra\ PSF at this location.
\label{opt}}
\end{minipage}
\end{center}

The spectrum of the compact source was extracted from a circular
region 2 pixels (0.984\arcsec) in radius. The main source of
background comes from thermal emission emitted by the gaseous remnant
itself, which accounts for half the extracted counts ($\sim$100 out of
193 total).  Our spectral fit for the compact star, therefore,
included a NEI thermal component with variable intensity, but fixed
abundance (0.3 times solar), temperature and ionization timescale. The
latter two quantities were determined by a spectral fit to data
extracted from an annulus surrounding the point source covering 4--6
pixels in radius.  Emission from the compact star was assumed to be
either a blackbody or a power-law. Absorption, including Galactic and
LMC components, was included as above.  Table~\ref{tab_star} shows the
results and the spectral data and model are plotted in
Fig.~\ref{spectrum} (in red). The red dashed curve shows the best fit
power-law model for the compact star's emission.  The luminosity of
the compact source under the power law spectral model is $L_X \sim
2-5\times 10^{33}$ erg s$^{-1}$ (0.5--2 keV) and $L_X \sim 3-6\times
10^{33}$ erg s$^{-1}$ (2--6 keV).  The blackbody fits yield similar
values: $L_X \sim 0.8-2\times 10^{33}$ erg s$^{-1}$ (0.5--2 keV) and
$L_X \sim 3-6\times 10^{33}$ erg s$^{-1}$ (2--6 keV).

Using FFT and epoch folding techniques we searched for a pulsed signal
from the compact object.  Nothing significant was found for
frequencies of several mHz up to the Nyquist limit ($\sim$0.154
Hz), which was set by the intrinsic time resolution ($\sim$3.24 s) of
our ACIS observation mode.  We are unable to set interesting limits on
the pulsed fraction because of the faintness of the source in
comparison to the contaminating remnant flux even within a
$\sim$1$^{\prime\prime}$ radius extraction region.

Radio images \citep{dick98} from the ATCA show no evidence for the
compact source in any of the four wavelengths mapped (3.5 cm, 6 cm, 13
cm, and 20 cm) to an estimated flux limit of $\sim$2 mJy.  These
limits are a factor of $\sim$20 below the radio flux of the pulsar
wind nebula in SNR 0453$-$685 \citep{gaensler03}, whereas the X-ray
source in \snr\ is only a factor of two fainter than the source in SNR
0453$-$685. Clearly \co\ is relatively radio quiet.  There is also no
clear optical counterpart in the DSS, but the field is crowded and the
flux limit is high. We obtained public domain images from the MACHO
Project that allowed us to set a more sensitive limit. In
Fig.~\ref{opt} we show a portion of an image made by stacking 10
individual 300-s exposures with moderately good seeing.  The combined
image was registered to the sky using several hundred HST guide stars
for a positional accuracy of $<$0.3$^{\prime\prime}$ in each axis.  A
subset of these stars was also used to estimate a coarse photometric
zero point for the image (accurate to roughly $\pm$0.3 mag).  The
nearest stars are 3.4$^{\prime\prime}$ (star A) and
1.1$^{\prime\prime}$ (star B) away from the X-ray compact object and
are therefore nominally outside the \chandra\ error circle as
indicated by the circle in the figure.  The estimated magnitudes of
these stars are $\sim$17 (A) and $\sim$19 (B).

\section{Conclusions and Summary}

There is a strong gradient in density from east to west across \snr\
as hinted at by the brightness gradient. The emitting gas density can
be determined from the fitted emission measure and a simple geometric
estimate for the volume. The east rim spectrum comes from a region
$\sim$10$^{\prime\prime}$ long by $\sim$3$^{\prime\prime}$ thick.
Assuming the depth in the line-of-sight is the same as its length, we
find a volume of $6.4\times 10^{55}$ cm$^3$, which results in a
hydrogen number density of 23 cm$^{-3}$.  The southeast rim spectrum
yields a similar density estimate of 10 cm$^{-3}$.  We estimate the
emission volume of the west rim spectrum (extracted from a region of
size $50^{\prime\prime}\times 15^{\prime\prime}$) to be
$\sim$$1.5\times10^{58}$ cm$^3$ which leads to a density estimate of 1
cm$^{-3}$.  There is more than a factor of 10 variation in the ambient
medium density from east to west. We can estimate ``shock'' ages from
the densities just derived and the ionization timescales in
Table~\ref{tab_snr}.  We find values of 55 yrs for the east rim and
500 yrs for the southeast rim, albeit with large errors. This suggests
that the SN shock wave has encountered some portions of the bright
eastern limb more recently than others. To conclude this discussion of
the remnant's environment we note that the open cluster HS114
\citep{hodsex66} lies on the high density side of the remnant; the
bright X-ray limb (near the east rim region) is only some
8$^{\prime\prime}$ (2 pc) from the edge of the cluster.

The complex morphology of \snr\ revealed by our new \chandra\ data
makes it clear that estimates of its dynamical state based on global
properties (e.g., size, brightness, integrated spectrum) are 
questionable.  We make the unjustified but plausible assumption
that the remnant's evolution toward the low density northwest
hemisphere can be described by an abiabatic Sedov-phase model.  The
radius of the remnant in this direction is $\sim$50$^{\prime\prime}$
(12 pc) and the ambient density, using the results in the previous
paragraph, is $\sim$0.25 cm$^{-3}$.  With these values and a canonical
explosion energy of $10^{51}$ ergs the age of \snr\ is $\sim$4600 yrs.

The high density environment, proximity to a young star cluster, and
presence of enhanced low-Z metal abundances in the remnant all point
toward a core collapse SN and massive star progenitor for \snr. A
neutron star (or, more speculatively, a black hole) is therefore a
plausible identification for the compact central X-ray star. The lack
of an extended nebula and the relatively low radio-to-X-ray flux ratio
argue against a young, rapidly-rotating, high spin-down-rate
pulsar. However the X-ray luminosity (a few times $10^{33}$ erg
s$^{-1}$) and spectrum (e.g., $kT_{\rm BB} \sim 0.75-1.3$ keV) are broadly
consistent with the emerging class of central compact X-ray--emitting
objects such as the one in Cas A \citep{pav00} that are believed to be
isolated neutron stars. Unfortunately the current optical magnitude
limits cannot definitively rule out an extragalactic origin for the
X-ray source, although the star's location very near the center of
\snr\ gives strong support for a physical association. An unambiguous
identification of the X-ray source will require further study across
the electromagnetic spectrum.

\acknowledgments

This paper utilizes public domain data originally obtained by the
MACHO Project, whose work was performed under the joint auspices of
the U.S. Department of Energy, National Nuclear Security
Administration by the University of California, Lawrence Livermore
National Laboratory under contract No.\ W-7405-Eng-48, the National
Science Foundation through the Center for Particle Astrophysics of the
University of California under cooperative agreement AST-8809616, and
the Mount Stromlo and Siding Spring Observatory, part of the
Australian National University.  We also acknowledge J.~Dickel and
P.~Ghavamian, who helped with various aspects of this research.
Partial support was provided by \chandra\ grant no.~GO2-3069X to
Rutgers University, NASA grant no.~NAG5-9281 to SAO, and NASA Contract
NAS8-39073 to SAO.

\end{document}